\documentclass[aps,pre,twocolumn,groupedaddress,amsmath,eqsecnum,showpacs]{revtex4}

    \usepackage{bm}
    \usepackage{graphicx}
\newcommand\beq{\begin{equation}}
\newcommand\eeq{\end{equation}}
\newcommand\beqa{\begin{eqnarray}}
\newcommand\eeqa{\end{eqnarray}}
\newcommand{\nn}{\nonumber\\}
\newcommand{\hs}{\text{HS}}

\newcommand{\dd}{{d}}

\begin{document}
\title{Radial distribution function of penetrable sphere fluids to second order in density}
\author{Andr\'es Santos}
\email{andres@unex.es}
\homepage{http://www.unex.es/eweb/fisteor/andres/}
\affiliation{Departamento de F\'{\i}sica, Universidad de
Extremadura, E-06071 Badajoz, Spain}
\author{Alexandr Malijevsk\'y}
\email{malijevsky@icpf.cas.cz} \affiliation{E. H\'ala Laboratory of
Thermodynamics, Academy of Science of the Czech Republic, 16502
Prague 6, Czech Republic and Institute of Theoretical Physics,
Faculty of Mathematics and Physics, Charles University, 18000 Prague
8, Czech Republic}

\date{\today}
\begin{abstract}
The simplest bounded potential is that of  penetrable spheres, which
takes a positive finite value $\epsilon$ if the two spheres are
overlapped, being 0 otherwise. In this paper  we  derive the cavity
function  to second order in density and the fourth virial
coefficient as functions of $T^*\equiv k_BT/\epsilon$ (where $k_B$
is the Boltzmann constant and $T$ is the temperature) for penetrable
sphere fluids. The expressions are exact, except for the function
represented by an elementary diagram inside the core, which is
approximated by a polynomial form in excellent agreement with
accurate results obtained  by Monte Carlo integration. Comparison
with the hypernetted-chain (HNC) and Percus--Yevick (PY) theories
shows that the latter is better than the former for $T^*\lesssim 1$
only. However, even at zero temperature (hard sphere limit), the PY
solution is not accurate inside the overlapping region, where no
practical cancelation of the neglected diagrams  takes place. The
exact fourth virial coefficient is positive for $T^*\lesssim 0.73$,
reaches a minimum negative value at $T^*\approx 1.1$, and then goes
to zero from below as $1/{T^*}^4$ for high temperatures. These
features are captured qualitatively, but not quantitatively, by the
HNC and PY predictions. In addition, in both  theories the
compressibility route is the best one for $T^*\lesssim 0.7$, while
the virial route is preferable if $T^*\gtrsim 0.7$.

\end{abstract}
\maketitle

\section{Introduction\label{sec1}}
Ultrasoft and bounded potentials represent useful models to
characterize the effective two-body interaction in some colloidal
systems, such as star or chain polymers in good solvents
\cite{L01,LLWAJAR98,SS97,GL98,LLWL00,LBH00,LLWL01,FHL03}. The
simplest bounded potential is that of so-called penetrable spheres
(PS), which is defined as
\beq
\phi(r)=\left\{
\begin{array}{ll}
\epsilon,& r<\sigma,\\
0,&r>\sigma,
\end{array}
\right.
\label{1}
\eeq
where $\epsilon>0$.  This interaction potential was suggested by
Marquest and Witten \cite{MW89} as a simple theoretical approach to
the explanation of the experimentally observed crystallization of
copolymer mesophases and it  has been since then the subject of a
number of studies
\cite{LLWL01,KGRCM94,LWL98,S99,FLL00,RSWL00,SF02,KS02,CG03,AS04,S05,VS05,MS06}.
The classical integral equation theories, in particular the
Percus--Yevick (PY) and the hypernetted-chain (HNC) approximations,
do not describe satisfactorily well the structure of the PS fluid,
especially inside the overlapping region for low temperatures. Thus,
the PS model can be used as a stringent benchmark to test
alternative theories \cite{S99,FLL00,RSWL00,CG03,MS06}. {}From that
point of view, the derivation of exact properties provides an
invaluable tool. The exact structural and thermodynamic properties
of the PS fluid in the high-temperature limit $T^*\equiv
k_BT/\epsilon\to\infty$ (where $k_B$ is the Boltzmann constant and
$T$ is the temperature) are known for any density $\rho\sigma^3$,
including the high-density regime $\rho\sigma^3\sim T^*$
\cite{AS04}. On the other hand,  the corresponding properties in the
complementary low-density limit for any temperature has not been
addressed, to the best of our knowledge, except in the
one-dimensional case \cite{MS06}.

The aim of this paper is to derive the exact expressions for the
radial distribution function $g(r)$ and, equivalently,  the cavity
function $y(r)$ of PS fluids to second order in density. To that end
we will exploit the fact that the PS Mayer function is proportional
to the hard sphere (HS) Mayer function. This implies that the
diagrams to be evaluated are the same as in the case of HS, except
that now each diagram is affected by a temperature-dependent factor.

In the next Section we present some definitions and basic equations.
The density expansion  of $y(r)$ to second order is worked out in
Sec.\ \ref{sec3}, where the HS functions derived by Nijboer and van
Hove \cite{NvH52} outside the core $r>\sigma$ are complemented by
their extensions in the overlapping region ($r<\sigma$). However, we
have not been able to derive the rigorously exact expression for
$r<\sigma$ of the function $\chi(r)$ represented by the only
elementary diagram. Instead, the exact values of $\chi(0)$,
$\chi'(0)$, $\chi(\sigma)$, $\chi'(\sigma)$, $\chi''(\sigma)$,
$\chi'''(\sigma)$, and $\int_0^\sigma dr\, r^2\chi(r)$ are obtained
in Sec.\ \ref{sec4}. With these constraints, we have constructed a
polynomial approximation of $\chi(r)$ for $r<\sigma$ which yields
results indistinguishable from those obtained by Monte Carlo (MC)
integration with six significant figures. The exact fourth virial
coefficient is also derived in Sec.\ \ref{sec4}. The exact results
are compared with the HNC and PY predictions in Sec.\ \ref{sec5}. It
is seen that the latter is generally preferable at low temperatures,
while the former is more accurate at high temperatures. The paper
ends with the conclusion section.

\section{Definitions and basic equations\label{sec2}}
We consider in this paper a fluid  of  particles interacting via the
pairwise potential (\ref{1}). Henceforth we take $\sigma=1$ as the
length unit. Let us introduce the cavity (or background) function
\beq
y(r|\eta,T^*)=e^{\phi(r)/k_BT}g(r|\eta,T^*),
\label{2.1}
\eeq
where $g(r|\eta,T^*)$ is the radial distribution function,
$\eta\equiv (\pi/6)\rho$ being the packing fraction. Equation
(\ref{2.1}) implies that
\beq
g(r|\eta,T^*)=y(r|\eta,T^*)-x y(r|\eta,T^*)\Theta(1-r),
\label{2.17}
\eeq
where $\Theta(r)$ is the Heaviside step function and we have called
\beq
x\equiv 1-e^{-1/T^*}.
\label{2.18}
\eeq
The parameter $x$ represents the probability of rejecting an overlap
of two particles in a MC move. The thermodynamic quantities can be
expressed in terms of $g(r|\eta,T^*)$ or $y(r|\eta,T^*)$
\cite{B74,BH76,HM86}. Particularized to the PS model, the
compressibility factor $Z\equiv p/\rho k_BT$ is given by the virial
equation of state as
\beq
Z(\eta,T^*)=1+4\eta x y(1|\eta,T^*).
\label{n2.1}
\eeq
The (dimensionless) isothermal compressibility $K\equiv
k_BT\left({\partial \rho}/{\partial p}\right)_T$ is
\beqa
K(\eta,T^*)&=&1+24\eta \left\{\int_0^\infty dr\,
r^2\left[y(r|\eta,T^*)-1\right]\right.\nn &&\left.-x\int_0^1 dr\,
r^2 y(r|\eta,T^*)\right\}.
\label{n11}
\eeqa
Finally, the internal energy per particle can be written as
\beq
u(\eta,T^*)=\epsilon\left[\frac{3}{2}T^*+12\eta (1-x)\int_0^1 d r\,
r^2 y(r|\eta,T^*)\right].
\label{n2.3}
\eeq
These three quantities are thermodynamically connected by the
relations
\beq
K^{-1}=\frac{\partial(\eta Z)}{\partial\eta},
\label{n12}
\eeq
\beq
\eta \frac{\partial (u/\epsilon)}{\partial\eta}=(1-x)\frac{\partial
Z}{\partial x}.
\label{n13}
\eeq

The  series expansions  of the cavity function and the
compressibility factor in powers of density read
\beq
y(r|\eta,T^*)=1+\sum_{n=1}^\infty
y_n(r|T^*)\left(\frac{6}{\pi}\right)^n\eta^n,
\label{2.2}
\eeq
\beq
Z(\eta,T^*)=1+\sum_{n=1}^\infty b_{n+1}(T^*)\eta^n.
\label{n14}
\eeq
In Eq.\ \eqref{n14},  $b_{n}(T^*)$ is the (reduced) $n$th virial
coefficient. The quantities $\{b_n(T^*)\}$ can be obtained from the
functions $\{y_n(r|T^*)\}$ through the virial route, Eq.
\eqref{n2.1}, the compressibility route, Eq.\ \eqref{n11}, or the
energy route, Eq.\ \eqref{n2.3}. In order to distinguish the results
derived through each route, we will use the notation $b_{n}^v(T^*)$,
$b_{n}^c(T^*)$, $b_{n}^e(T^*)$, respectively. Of course,
$b_{n}^v(T^*)=b_{n}^c(T^*)=b_{n}^e(T^*)$ if the exact cavity
function is employed.

 Insertion of the expansion \eqref{2.2} into Eqs.\
\eqref{n2.1} and \eqref{n2.3} yields (for $n\geq 2$)
\beq
b_n^v(T^*)=4x\left(\frac{6}{\pi}\right)^{n-2}y_{n-2}(1|T^*),
\label{n15}
\eeq
\beq
b_n^e(T^*)=12(n-1)\left(\frac{6}{\pi}\right)^{n-2}\int_0^xdx_1\int_0^1dr\,r^2y_{n-2}(r|T_1^*).
\label{n16}
\eeq
In Eq.\ \eqref{n16} use has been made of Eq.\ \eqref{n13} and of the
ideal gas condition $\lim_{T^*\to\infty} b_n(T^*)=0$.  In the case
of the compressibility route, insertion of Eq.\ \eqref{2.2} into
Eq.\ \eqref{n11} and use of the relation \eqref{n12} leads to the
recursive formula
\beq
b_n^c(T^*)=-\sum_{m=1}^{n-1}\frac{m}{n}b_m^c(T^*)K_{n-m}(T^*),
\label{n17}
\eeq
where $K_1(T^*)=-8x$ and
\beqa
K_n(T^*)&\equiv &24
\left(\frac{6}{\pi}\right)^{n-1}\left[\int_0^\infty dr\, r^2
y_{n-1}(r|T^*)\right. \nn &&\left.-x\int_0^1dr\, r^2
y_{n-1}(r|T^*)\right]
\label{n18}
\eeqa
for $n\geq 2$.

\section{Cavity function to second order in density \label{sec3}}
The virial coefficients $y_n(r|T^*)$
 are represented by diagrams \cite{B74,HM86}. In particular,
\beq
y_1(r|T^*)=
\begin{picture}(40,40)(-5,5)
\setlength{\unitlength}{.1mm}
\put(50,100){\circle*{10}}
\put(100,0){\circle{10}}
\put(0,0){\circle{10}}
\put(50,100){\line(1,-2){48}}
\put(50,100){\line(-1,-2){48}}
\end{picture},
\label{2.3}
\eeq
\beqa
y_2(r|T^*)&=&
\begin{picture}(40,40)(-5,5)
\setlength{\unitlength}{.1mm}
\put(0,100){\circle*{10}}
\put(100,100){\circle*{10}}
\put(100,0){\circle{10}}
\put(0,0){\circle{10}}
\put(0,100){\line(100,0){100}}
\put(0,100){\line(0,-100){95}}
\put(100,100){\line(0,-100){95}}
\end{picture}
 +2
\begin{picture}(40,40)(-5,5)
\setlength{\unitlength}{.1mm}
\put(0,100){\circle*{10}}
\put(100,100){\circle*{10}}
\put(100,0){\circle{10}}
\put(0,0){\circle{10}}
\put(0,100){\line(100,0){100}}
\put(0,100){\line(0,-100){95}}
\put(100,100){\line(0,-100){95}}
\put(100,100){\line(-1,-1){95}}
\end{picture}+
\frac{1}{2}
\begin{picture}(40,40)(-5,5)
\setlength{\unitlength}{.1mm}
\put(0,100){\circle*{10}}
\put(100,100){\circle*{10}}
\put(100,0){\circle{10}}
\put(0,0){\circle{10}}
\put(0,100){\line(0,-100){95}}
\put(100,100){\line(0,-100){95}}
\put(100,100){\line(-1,-1){95}}
\put(0,100){\line(1,-1){95}}
\end{picture}
\nn &&
 + \frac{1}{2}
\begin{picture}(40,40)(-5,5)
\setlength{\unitlength}{.1mm}
\put(0,100){\circle*{10}}
\put(100,100){\circle*{10}}
\put(100,0){\circle{10}}
\put(0,0){\circle{10}}
\put(0,100){\line(100,0){100}}
\put(0,100){\line(0,-100){95}}
\put(100,100){\line(0,-100){95}}
\put(100,100){\line(-1,-1){95}}
\put(0,100){\line(1,-1){95}}
\end{picture}.
\label{2.4}
\eeqa
Here, the open circles represent \textit{root} points separated by a
distance $r$, the filled circles represent \textit{field} points to
be integrated out, and each bond represents a Mayer function
\beq
f(r|T^*)=e^{-\phi(r)/k_BT}-1.
\label{2.5}
\eeq
Thus, for instance,
\beq
\begin{picture}(40,40)(-5,5)
\setlength{\unitlength}{.1mm}
\put(50,100){\circle*{10}}
\put(25,90){3}
\put(100,0){\circle{10}}
\put(110,-10){2}
\put(0,0){\circle{10}}
\put(-25,-10){1}
\put(50,100){\line(1,-2){48}}
\put(50,100){\line(-1,-2){48}}
\end{picture}
=\int \dd\mathbf{r}_3 \,f(r_{13}|T^*)f(r_{23}|T^*),
\label{2.6}
\eeq
\beqa
\begin{picture}(40,40)(-5,5)
\setlength{\unitlength}{.1mm}
\put(0,100){\circle*{10}}
\put(-25,90){3} \put(100,100){\circle*{10}}
\put(110,90){4}
\put(100,0){\circle{10}}
\put(110,-10){2}
\put(0,0){\circle{10}}
\put(-25,-10){1}
\put(0,100){\line(100,0){100}}
\put(0,100){\line(0,-100){95}}
\put(100,100){\line(0,-100){95}}
\put(100,100){\line(-1,-1){95}}
\end{picture}&=&
\int \dd\mathbf{r}_3\int \dd\mathbf{r}_4 \,
f(r_{13}|T^*)f(r_{34}|T^*)\nn &&\times f(r_{24}|T^*)f(r_{14}|T^*),
\label{2.7}
\eeqa
where $r_{ij}=|\mathbf{r}_i-\mathbf{r}_j|$ and $r_{12}=r$.

Equations (\ref{2.3})--(\ref{2.7}) hold for any interaction
potential. In the special case of PS,   the Mayer function becomes
\beq
f(r|T^*)=x f_{\text{HS}}(r),
\label{2.8}
\eeq
where
\beq
f_{\text{HS}}(r)= - \Theta(1-r)
\label{2.9}
\eeq
is the Mayer function of HS. Therefore, the spatial dependence of
each one of the diagrams contributing to the virial expansion
(\ref{2.2}) is exactly the same as for HS. The only difference is
that each diagram is now multiplied by the temperature-dependent
parameter $x$ raised to a power equal to the number of bonds in that
particular diagram. As a consequence, Eqs.\ \eqref{2.3} and
\eqref{2.4} become
\beq
y_1(r|T^*)=x^2 \gamma(r),
\label{n1}
\eeq
\beq
y_2(r|T^*)=x^3 \varphi(r)+2x^4
\psi(r)+\frac{x^4}{2}\gamma^2(r)+\frac{x^5}{2}\chi(r).
\label{n2}
\eeq
Here, $\gamma(r)$ is represented by the diagram on the right-hand
side of Eq.\ \eqref{2.3}, except that now each bond corresponds to a
Mayer function $f_\hs$ . Analogously, the functions $\varphi(r)$,
$\psi(r)$, and $\chi(r)$ are represented by the first, second, and
fourth diagram, respectively, on the right-hand side of Eq.\
\eqref{2.4}, with $f_\hs$ for each bond. The expressions of these
functions for $r>1$ are known \cite{NvH52,LGKM03}. The region $r>1$
is the physically relevant one in the case of HS. However, the
overlapping region $r<1$ is essential in the case of PS, since
$g(r|\eta,T^*)\neq 0$ for $r<1$, except in the zero-temperature
limit (where the PS model reduces to the HS one). Therefore, it is
necessary to extend the knowledge of $\gamma(r)$, $\varphi(r)$,
$\psi(r)$, and $\chi(r)$ to the domain $0\leq r\leq 1$.

Given a radial function $F(r)$ we define its Fourier transform as
\beq
\widetilde{F}(k)=\int d\mathbf{r}\,
e^{i\mathbf{k}\cdot\mathbf{r}}F(r)=\frac{4\pi}{k}\int_0^\infty dr\,
r \sin(kr)F(r).
\label{Fou}
\eeq
It is easy to realize that
$\widetilde{\gamma}(k)=[\widetilde{f}_\hs(k)]^2$, where
\beq
\widetilde{f}_\hs(k)=4\pi\frac{k\cos k-\sin k}{k^3}.
\label{n3}
\eeq
Inverse Fourier transform simply yields
\beq
\gamma(r)=\frac{\pi}{12}(2-r)^2(r+4)\Theta(2-r).
\label{n4}
\eeq
This implies that the function $\gamma(r)$ for $0\leq r\leq 1$ is
just the analytical continuation of its expression for $1\leq r\leq
2$. Next, note that
$\widetilde{\varphi}(k)=[\widetilde{f}_\hs(k)]^3$, so that
\beq
\varphi(r)=\varphi_A(r)\Theta(1-r)+\varphi_B(r)\Theta(3-r)
\label{n5}
\eeq
with
\beq
\varphi_A(r)=\frac{\pi^2}{36}\frac{3}{35r}(r-1)^4(r^3+4r^2-53r-162),
\label{n6}
\eeq
\beq
\varphi_B(r)=-\frac{\pi^2}{36}\frac{1}{35r}(r-3)^4(r^3+12r^2+27r-6).
\label{n7}
\eeq
Therefore, $\varphi(r)=\varphi_A(r)+\varphi_B(r)$ for $0\leq r\leq
1$, while $\varphi(r)=\varphi_B(r)$ for $1\leq r\leq 3$. In the case
of $\psi(r)$, one has
$\widetilde{\psi}(k)=\widetilde{f}_\hs(k)\widetilde{\gamma}^*(k)$,
where $\gamma^*(r)=\gamma(r)f_\hs(r)$. As a consequence,
\beq
\psi(r)=\psi_A(r)\Theta(1-r)+\psi_B(r)\Theta(2-r)
\label{n8}
\eeq
with
\beq
\psi_A(r)=-\frac{2}{3}\varphi_A(r),
\label{n9}
\eeq
\beq
\psi_B(r)=\frac{\pi^2}{36}\frac{1}{35r}(r-2)^2(r^5+4r^4-51r^3-10r^2+479
r-81).
\label{n10}
\eeq

Equations (\ref{n7}) and \eqref{n10} coincide with those derived in
Ref.\ \cite{NvH52} by a different method. On the other hand, the
functions $\varphi_A(r)$ and $\psi_A(r)$, which are needed to get
$\varphi(r<1)$ and $\psi(r<1)$, respectively, were not considered in
Refs.\ \cite{NvH52} and \cite{LGKM03}.  Near the origin,
\beq
\gamma(r)=\frac{\pi}{6}\left(8-6r\right)+\mathcal{O}(r^2),
\label{n21}
\eeq
\beq
\varphi(r)=-\frac{\pi^2}{36}30+\mathcal{O}(r^2),
\label{n22}
\eeq
\beq
\psi(r)=\frac{\pi^2}{36}\left(30-15r\right)+\mathcal{O}(r^2).
\label{n23}
\eeq
Equations\ \eqref{n5}--\eqref{n10} show that $\varphi(r)$ has a
fourth-order discontinuity at $r=1$ and at $r=3$, while $\psi(r)$
has a fourth-order discontinuity at $r=1$ and a second-order
discontinuity at $r=2$.

Now we turn to the much more involved  function $\chi(r)$,
represented by the elementary  diagram at the end of the right-hand
side of Eq.\ \eqref{2.4}. Let us decompose it in a form similar to
Eqs.\ \eqref{n5} and \eqref{n8},
\beq
\chi(r)=\chi_A(r)\Theta(1-r)+\chi_B(r)\Theta(\sqrt{3}-r)-\gamma^2(r).
\label{n19}
\eeq
The exact expression for $\chi_B(r)$ was obtained by Nijboer and van
Hove \cite{NvH52}. It reads
\beqa
\chi_B(r)&=&\pi\left[-r^2\left(\frac{3r^2}{280}-\frac{41}{420}\right)\sqrt{3-r^2}
-\left(\frac{23}{15}r-\frac{36}{35r}\right)
  \right.\nn
&&\times\cos^{-1}\frac{r}{\sqrt{3(4-r^2)}}    +\left(\frac{3
r^6}{560}-\frac{r^4}{15}+\frac{r^2}{2}+\frac{2r}{15}\right.\nn
&&\left.-\frac{9}{35r}\right)
\cos^{-1}\frac{r^2+r-3}{\sqrt{3(4-r^2)}} +\left(\frac{3
r^6}{560}-\frac{r^4}{15}\right.\nn &&\left.\left.
+\frac{r^2}{2}-\frac{2r}{15}+\frac{9}{35r}\right)
\cos^{-1}\frac{-r^2+r+3}{\sqrt{3(4-r^2)}}\right].
\label{n20}
\eeqa

We have not been able to obtain an analytic expression for $\chi(r)$
in the interval $0\leq r\leq 1$. By working with bipolar
coordinates, it is possible to express the derivative $\chi'(r<1)$
as a sum of 13 triple integrals, but only two of them seem to be
analytically solvable. Therefore, we have resorted to numerical
evaluation of $\chi(r<1)$ by the MC method \cite{LGKM03} and to a
very accurate polynomial approximation. In order to construct the
latter, some constraints on the exact $\chi(r<1)$ are derived in the
next Section.

\section{Constraints on $\chi(r)$. Polynomial
approximation \label{sec4}}

In this Section we derive some constraints on $\chi(r)$ for $r\leq
1$. First, we take into account that $\chi(r)$ and its first three
derivatives must be continuous at $r=1$. We are not aware of a
formal proof of this  statement, but it is strongly supported by the
following two arguments: (i) both $\varphi(r)$ and $\psi(r)$ have a
fourth-order discontinuity at $r=1$, even though a diagonal bond is
added when going from the diagram representing $\varphi(r)$ to that
representing $\psi(r)$; (ii) in the one-dimensional case, the three
functions $\varphi(r)$, $\psi(r)$, and $\chi(r)$  have the same type
of singularity at $r=1$, namely a second-order discontinuity
\cite{MS06}.

{}From Eqs.\ \eqref{n19} and \eqref{n20} one can get
\beq
\chi(1)=
\frac{\pi^2}{36}\left(\frac{b_4^\hs}{2}-\frac{57}{4}\right),
\label{4.2}
\eeq
\beq
\chi'(1)=
\frac{\pi^2}{36}\frac{1}{51}\left(\frac{347b_4^\hs}{3}-\frac{7219}{6}-\frac{256\sqrt{2}}{\pi}\right),
\label{4.3}
\eeq
\beq
\chi''(1)=
\frac{\pi^2}{36}\frac{1}{153}\left(-\frac{619b_4^\hs}{3}-\frac{8149}{6}+\frac{2432\sqrt{2}}{\pi}\right),
\label{4.4}
\eeq
\beq
\chi'''(1)=
\frac{\pi^2}{36}\frac{2}{153}\left(\frac{946b_4^\hs}{3}-\frac{16597}{3}-\frac{4832\sqrt{2}}{\pi}\right).
\label{4.5}
\eeq
In the above equations,
\beq
b_4^\hs=\frac{2707}{70}+\frac{438\sqrt{2}-4131 \sec^{-1}
3}{70\pi}\simeq 18.3648
\label{4.6}
\eeq
is the exact value of the fourth virial coefficient for HS.

Next, note that
\beqa
\chi(0)&=&\int d\mathbf{r}\,
\gamma(r)f_\hs(r)=-\frac{\pi^2}{3}\int_0^1dr\, r^2(r-2)^2(r+4)\nn
&=&-\frac{\pi^2}{36}30.
\label{4.1}
\eeqa
The same result is obtained from the following zero-separation
theorem for HS \cite{L95}:
\beq
\ln y_\hs(0|\eta)=4\eta y_\hs(1|\eta)+4\int_0^\eta d\eta_1
y_\hs(1|\eta_1),
\label{x36}
\eeq
where $y_\hs(r|\eta)=\lim_{T^*\to0}y(r|\eta,T^*)$ is the cavity
function for HS. As a further constraint on the unknown function
$\chi(r)$ for $r<1$, let us consider the alternative zero-separation
theorem
\beq
\frac{y_\hs'(0|\eta)}{y_\hs(0|\eta)}=-6\eta y_\hs(1|\eta).
\label{x37}
\eeq
This implies that $\lim_{T^*\to 0} y_2'(0|T^*)=-(\pi^2/36)63$.
{}From Eqs.\ \eqref{n2} and \eqref{n21}--\eqref{n23} one then has
\beq
\chi'(0)=\frac{\pi^2}{36}30.
\label{4.13}
\eeq
 As a consequence of Eqs.\
\eqref{n21}--\eqref{n23}, \eqref{4.1}, and \eqref{4.13}, the form of
$y_2$ near the origin is
\beqa
y_2(r|T^*)&=&\frac{\pi^2}{36}
x^3\left[-30+2x(46-39r)-15x^2(1-r)\right]\nn &&+\mathcal{O}(r^2).
\label{4.14}
\eeqa

Let us apply now the condition of thermodynamic consistency for the
fourth virial coefficient $b_4(T^*)$. Taking into account that
$y_0(r|T^*)=1$, $y_1(1|T^*)=x^2\gamma(1)=(5\pi/12)x^2$, and
\beqa
y_2(1|T^*)&=&x^3\varphi(1)+2x^4\psi(1)+\frac{x^4}{2}\gamma^2(1)+\frac{x^5}{2}\chi(1)\nn
&=&\frac{\pi^2}{36}x^3\left[-\frac{544}{35}+\frac{6347}{280}x+\left(\frac{b_4^\hs}{4}-\frac{57}{8}\right)x^2\right],\nn
&&
\label{4.7}
\eeqa
Eq.\ \eqref{n15} yields
\beq
b_2(T^*)=4x,\quad b_3(T^*)=10x^3,
\label{4.8}
\eeq
\beq
b_4(T^*)=x^4\left[b_4^\hs x^2-(1-x)\frac{4352-1995x}{70}\right].
\label{4.9}
\eeq
The same results for $b_2$ and $b_3$ are obtained through the energy
route, Eq.\ \eqref{n16}. As for $b_4$, Eq.\ \eqref{n16} yields
\beqa
b_4(T^*)&=&\frac{36}{\pi^2} 9x^4\int_0^1
dr\,r^2\left[\varphi(r)+\frac{8}{5}x\psi(r)\right.\nn
&&\left.+\frac{2}{5}x\gamma^2(r)+\frac{1}{3}x^2\chi(r)\right].
\label{4.10}
\eeqa
Since the functions $\varphi(r)$, $\psi(r)$, and $\gamma^2(r)$ are
known for $r<1$, the integrals involving them can be performed.
Thus, equating the right-hand sides of Eqs.\ \eqref{4.9} and
\eqref{4.10} one gets
\beq
\int_0^1dr\,
r^2\chi(r)=\frac{\pi^2}{36}\left(\frac{b_4^\hs}{3}-\frac{57}{6}\right)=\frac{2}{3}\chi(1).
\label{4.11}
\eeq
In turn, this condition implies that
\beq
K_3=-{2}x^3\left[256-\frac{12752}{35}x+\frac{6347}{35}x^2+\left(2b_4^\hs-57\right)x^3\right],
\label{4.12}
\eeq
where the coefficients $K_n$ are defined by Eq.\ \eqref{n18}. Use of
$K_1=-8x$, $K_2=2x^2(32-15x)$, and \eqref{4.12} in Eq.\ \eqref{n17}
leads again to Eqs.\ \eqref{4.8} and \eqref{4.9} \cite{H57}.

Although the exact analytic expression of $\chi_A(r)$, and hence of
$\chi(r)$ for $r<1$ is not known, we have derived in this Section a
number of constraints. The value of $\chi(r)$ and its  first three
derivatives at $r=1$ are given by Eqs.\ \eqref{4.2}--\eqref{4.5}. On
the other hand, Eqs.\ \eqref{4.1} and \eqref{4.13} give $\chi(r)$
and $\chi'(r)$ at the origin. Finally, the integral of $r^2\chi(r)$
in the interval $0\leq r\leq 1$ is determined by Eq.\ \eqref{4.11}.
Since there are seven constraints we can approximate $\chi(r)$ for
$r\leq 1$ by a polynomial of sixth degree:
\beqa
\chi_{\text{poly}}(r)&=&\frac{\pi^2}{36}\left[\alpha_0+\alpha_1(r-1)+\alpha_2(r-1)^2\right.\nn
&&\left.+\alpha_3(r-1)^3+(r-1)^4\left(\beta_0+\beta_1 r+\beta_2
r^2\right)\right].\nn
\label{4.15}
\eeqa
In this equation, the constants $\alpha_0=(36/\pi^2)\chi(1)$,
$\alpha_1=(36/\pi^2)\chi'(1)$, $\alpha_2=(36/\pi^2)\chi''(1)/2$, and
$\alpha_3=(36/\pi^2)\chi'''(1)/6$ are obtained from Eqs.\
\eqref{4.2}--\eqref{4.5}. {}From Eqs.\ \eqref{4.1} and \eqref{4.13}
one gets
\beq
\beta_0=\frac{1}{459}\left(\frac{4309b_4^\hs}{3}-\frac{129317}{6}-\frac{10784\sqrt{2}}{\pi}\right),
\label{4.16}
\eeq
\beq
\beta_1=\frac{1}{27}\left(\frac{554b_4^\hs}{3}-\frac{8663}{3}-\frac{1120\sqrt{2}}{\pi}\right).
\label{4.17}
\eeq
Finally, application of \eqref{4.11} yields
\beq
\beta_2=\frac{1}{3}\left(\frac{3803b_4^\hs}{6}-\frac{134713}{12}-\frac{920\sqrt{2}}{\pi}\right).
\label{4.18}
\eeq
The second derivative at the origin is
\beqa
\chi_{\text{poly}}''(0)&=&\frac{\pi^2}{36}\frac{10}{459}\left(\frac{55069b_4^\hs}{3}-\frac{981592}{3}-
\frac{22232\sqrt{2}}{\pi}\right)\nn &\simeq &-\frac{\pi^2}{36}
2.07929.
\label{4.19}
\eeqa
In contrast, the exact result is (see the Appendix)
\beq
\chi''(0)=-\frac{\pi^2}{36}\left(12-\frac{18\sqrt{3}}{\pi}\right)\simeq
-\frac{\pi^2}{36}2.07608.
\label{4.20}
\eeq
Therefore, $\chi_{\text{poly}}''(0)/\chi''(0)\simeq 1.00155$. This
gives an idea of the extreme accuracy of $\chi_{\text{poly}}(r)$. In
fact, we have evaluated numerically $\chi(r)$ by MC integration with
6 significant figures and have found that $\chi_{\text{poly}}(r)$
agrees with  $\chi(r)$ within the error bars (see Table
\ref{table1}). One could exploit the exact knowledge of $\chi''(0)$,
Eq.\ \eqref{4.20}, to propose an approximation presumably even more
accurate than Eq.\ \eqref{4.15}, but this does not seem to be
necessary in view of Table \ref{table1}.
\begin{table}
\caption{\label{table1} Values of $-(36/ \pi^2)\chi(r)$ in the
interval $0\leq r\leq 1$ as obtained numerically by MC integration
and as given by the polynomial approximation \protect\eqref{4.15}.
The number enclosed between parentheses in the second column
indicates the 95\%-confidence error.}
\begin{ruledtabular}
\begin{tabular}{ccc}
$r$&MC &Eq.\ \protect\eqref{4.15}
\\
\hline
0.00  & 29.9994(6)& 30\\
0.05 &       28.5029(5) &         28.5031\\
0.10 &  27.0146(5) &  27.0144\\
0.15 &  25.5367(4)&  25.5369\\
0.20& 24.0735(4)& 24.0736\\
0.25& 22.6277(4)&  22.6275\\
0.30 & 21.2015(3) &  21.2017\\
0.35 & 19.7987(3) &  19.7991\\
0.40& 18.4228(3) &   18.4228\\
0.45&  17.0756(3)&17.0758\\
 0.50 &  15.7614(2) &15.7612\\
0.55  & 14.4822(2) & 14.4820\\
0.60 &  13.2414(2) & 13.2413\\
0.65 &  12.0420(2) & 12.0421\\
0.70 &  10.88745(13) &   10.88740\\
0.75 &  9.78032(12)& 9.78037\\
0.80 & 8.72413(14)& 8.72404\\
0.85 &  7.72152(10)& 7.72151\\
0.90&  6.77582(8)&  6.77586\\
0.95  &  5.89018(7) & 5.89019\\
1.00&  5.06759(5)&  5.06762
\end{tabular}
\end{ruledtabular}
\end{table}

\section{Comparison with the HNC and PY theories\label{sec5}}
Once we have obtained the exact temperature-dependence of the
function $y_2(r)$ and  the associated fourth virial coefficient
$b_4$ for PS, it is worthwhile comparing these two quantities with
the predictions provided by the two classical integral equation
theories, namely the HNC and PY theories.

\subsection{Cavity function to second order, $y_2(r)$}
 In the HNC theory,
the elementary diagrams are neglected at any order in density
\cite{HM86}. To second order in density, the only elementary diagram
is the last one given in Eq.\ \eqref{2.4}. Therefore,  the function
$y_2(r)$ is approximated by
\beq
y_2^{\text{HNC}}(r|T^*)=x^3 \varphi(r)+2x^4
\psi(r)+\frac{x^4}{2}\gamma^2(r).
\label{HNC}
\eeq
 In the PY
approximation, apart from the elementary diagrams, a subset of the
remaining diagrams is also neglected. In particular, the PY
expression for $y_2(r)$ only retains the two first diagrams in Eq.\
\eqref{2.4}, so that
\beq
y_2^{\text{PY}}(r|T^*)=x^3 \varphi(r)+2x^4 \psi(r).
\label{PY}
\eeq

\begin{figure}[htb]
\includegraphics[width=\columnwidth]{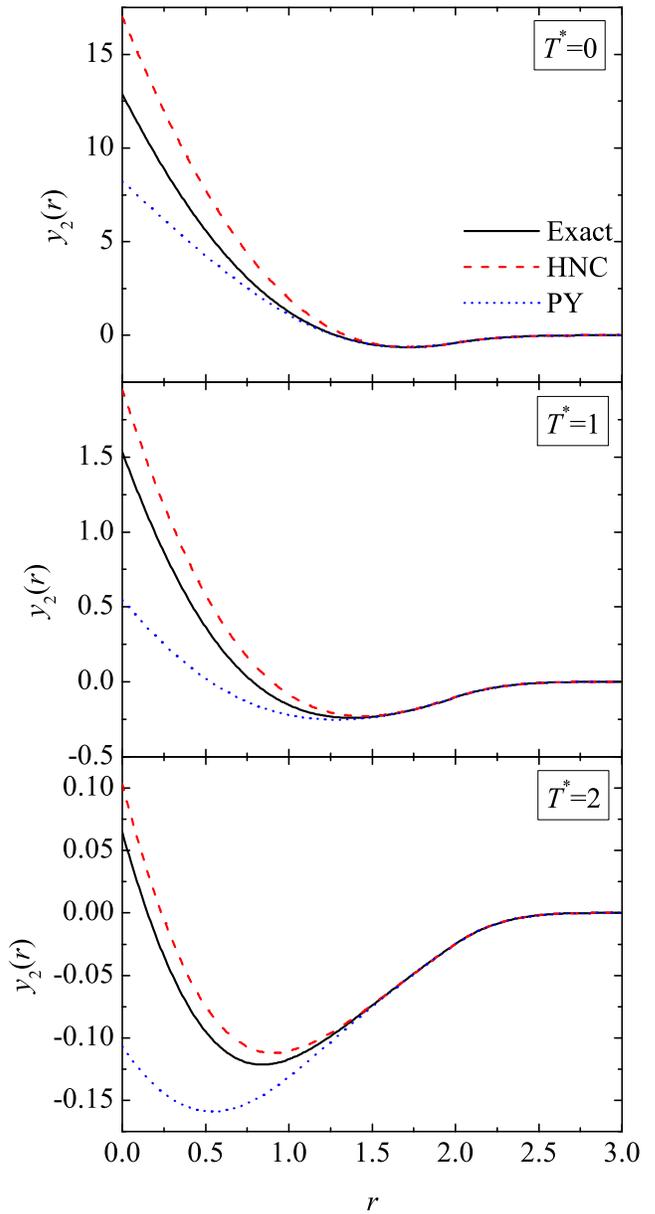}
\caption{(Color online) Plot of the function $y_2(r)$ at $T^*=0$
(top panel), $T^*=1$ (middle panel), and $T^*=2$ (bottom panel). The
solid lines are the exact results, the dashed lines are the HNC
predictions, and the dotted lines are the PY predictions.
\label{y2}}
\end{figure}

Figure \ref{y2} compares the exact function $y_2(r)$ with the HNC
and PY approximations at $T^*=0$ (hard spheres), $T^*=1$, and
$T^*=2$. Both theories agree very well with the exact $y_2(r)$ for
$r\geq 1.5$ but discrepancies are apparent for shorter distances,
especially inside the core ($r<1$). Although restricted to low
densities, Fig.\ \ref{y2} clearly illustrates some of the general
features found  at finite densities \cite{FLL00,RSWL00}: the HNC
overestimates the penetrability effect, while the PY approximation
underestimates it. The former property is a consequence of the
neglect of $(x^5/2)\chi(r)$, which is a negative definite quantity.
This is only partially compensated by the PY neglect of
$(x^4/2)\gamma^2(r)$, since $\gamma^2(r)>|\chi(r)|$ for $r<1$ and,
moreover, $x^4\geq x^5$. While the PY theory tends to be better at
lower temperatures (i.e., when the overlapping of particles is
hindered and the system is close to that of HS), the HNC is
preferable at higher temperatures. If we characterize the quality of
each approximation by the separation of the corresponding contact
value $y_2(1)$ from the exact result, it turns out that the
temperature beyond which the HNC approximation becomes better than
the PY approximation is $T^* \simeq 1.04$. This is similar to the
behavior found in the one-dimensional case \cite{MS06}.

\subsection{Fourth virial coefficient}
\begin{table*}[htb]
\caption{\label{table2}Fourth virial coefficient $b_4(T^*)$ and
other related quantities as given exactly and by the HNC and PY
approximations through the virial (v), energy (e), and
compressibility (c) routes.}
\begin{ruledtabular}
\begin{tabular}{cccccc}
Theory &$b_4(T^*)$&$b_4(0)$&$T_0^*$&$T^*_{\text{min}}$&$\left.b_4\right|_{\text{min}}$\\
 \hline
Exact&$x^4[b_4^\hs x^2-(1-x)({4352-1995
 x})/{70}]$&$b_4^\hs$&$0.7250$&$1.1027$&$-1.4803$\\
HNC,v/e
&$x^4(6347x-4352)/{70}$&$\frac{57}{2}$&$0.8641$&$1.2574$&$-1.1258$\\
HNC,c&$x^4(31735x-26112)/{420}$&$\frac{5623}{420}$&$0.5778$&$0.9314$&$-2.3345$\\
PY,v &${16}x^4(171x-136)/{35}$&$16$&$0.6304$&$0.9888$&$-2.0378$\\
PY,e &${2}x^4(6347x-5440)/{175}$&$\frac{1814}{175}$&$0.5140$&$0.8641$&$-2.7485$\\
PY,c&$x^4(6347x-4352)/{105}$&$19$&$0.8641$&$1.2574$&$-0.7505$
\end{tabular}
\end{ruledtabular}
\end{table*}
The knowledge of $y_2^{\text{HNC}}(r|T^*)$ and
$y_2^{\text{PY}}(r|T^*)$ allows one to get the associated
expressions for the fourth virial coefficient $b_4(T^*)$. As
discussed in Section \ref{sec2}, there are three alternative routes
[cf. Eqs.\ \eqref{n15}--\eqref{n17}] and there is no reason to
expect internal consistency among them, unless the exact $y_2(r)$ is
used. The expressions for $b_4^v(T^*)$, $b_4^e(T^*)$, and
$b_4^c(T^*)$ in the HNC and PY approximations are given in Table
\ref{table2}, where, for completeness, the exact  expression, Eq.\
\eqref{4.9}, is also included. It is known \cite{nBH76} that the HNC
integral equation provides thermodynamically consistent results
through the virial and energy routes, regardless of the potential
considered. This explains the fact that
$b_4^{\text{HNC},v}(T^*)=b_4^{\text{HNC},e}(T^*)$. On the other
hand, the PY integral equation yields three different predictions,
i.e., $b_4^{\text{PY},v}(T^*)\neq b_4^{\text{PY},e}(T^*)\neq
b_4^{\text{PY},c}(T^*)$. It is interesting to note that
$b_4^{\text{HNC},v/e}(T^*)=\frac{3}{2}b_4^{\text{PY},c}(T^*)=\frac{1}{4}xd
b_4^{\text{PY},e}(T^*)/dx$.

In the limit $T^*\to 0$ one recovers the  known results for HS,
namely $b_4^{\text{HNC},v/e}(0)=\frac{57}{2}=28.5$,
$b_4^{\text{HNC},c}(0)=\frac{5623}{420}\simeq 13.3881$,
$b_4^{\text{PY},v}(0)=16$, and $b_4^{\text{PY},c}(0)=19$. Although
the energy route is ill defined for strict HS, taking the
zero-temperature limit on the PS model yields well defined values
\cite{S06}. In that way, one finds
$b_4^{\text{PY},v}(0)=\frac{1814}{175}\simeq 10.3657$, which is a
rather poor value reflecting the inaccuracy at any temperature of
$y_2^\text{PY}(r)$ for $r<1$.  In the opposite high-temperature
limit, one has $\lim_{T^*\to\infty} b_4(T^*)/x^4=-2176/35$. This
exact value is retained by all the approximations, except by the
compressibility route in the PY theory, which yields
$\lim_{T^*\to\infty} b_4^{\text{PY},c}(T^*)/x^4=-4352/105$, i.e.,
$2/3$ of the exact result.

While $b_2(T^*)$ and $b_3(T^*)$ are positive definite quantities,
this is not the case of $b_4(T^*)$. The latter quantity  changes
sign at a certain ``Boyle-like'' temperature $T_0^*$. In addition,
$b_4(T^*)$ presents a (negative) minimum value
$\left.b_4\right|_{\text{min}}$ at a temperature
$T^*_{\text{min}}>T_0^*$. The numerical values of $T_0^*$,
$T^*_{\text{min}}$, and $\left.b_4\right|_{\text{min}}$ are
displayed in Table \ref{table2}. {}From Eq.\ \eqref{n16} one can see
that the temperature $T^*_{\text{min}}$ associated with $b_4^e(T^*)$
is the temperature across which the integral $\int_0^1dr\, r^2
y_2(r|T^*)$, and hence the third-order term in the density expansion
of the internal energy, changes from positive to negative.

\begin{figure}[htb]
\includegraphics[width=\columnwidth]{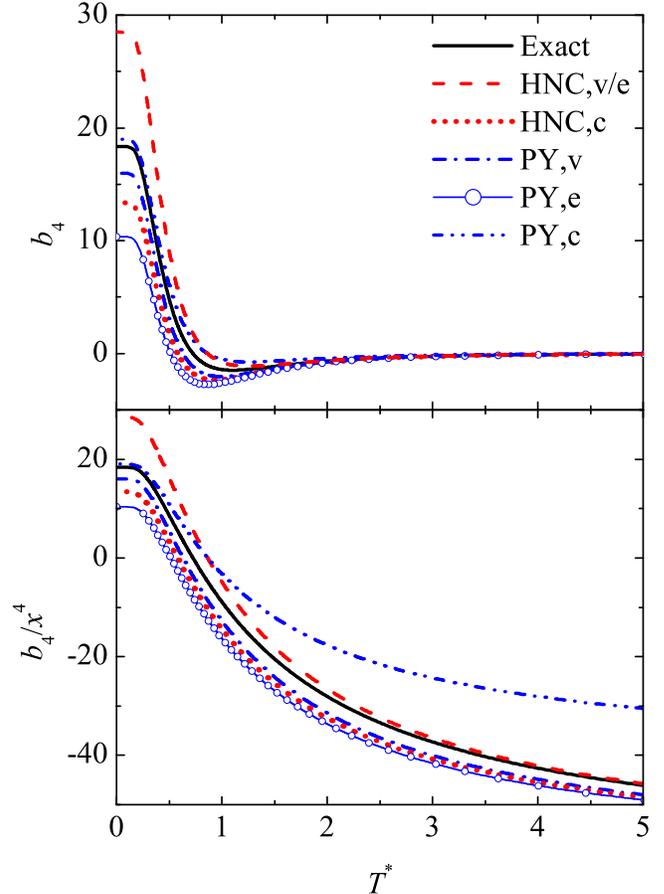}
\caption{(Color online) Plot of the fourth virial coefficient
$b_4(T^*)$ (top panel) and of the scaled fourth virial coefficient
$b_4(T^*)/x^4$ (bottom panel), where $x=1-e^{-1/T^*}$,  as given
exactly and by the HNC and PY approximations.
\label{b4}}
\end{figure}
The temperature dependence of the fourth virial coefficient  is
shown in Fig.\ \ref{b4}, where, apart from the exact curves, the two
HNC approximations and the three PY approximations are included. The
best approximation up to $T^*\simeq 0.71$ is provided by
$b_4^{\text{PY},c}$. In the intermediate range $0.71\lesssim
T^*\lesssim 1.04$, however, $b_4^{\text{PY},v}$ presents the best
agreement. Finally, for $T^*\gtrsim 1.04$ the best performance
corresponds to $b_4^{\text{HNC},v/e}$. Within the PY theory, the
energy route is never better than the virial route but becomes
preferable to the compressibility route for $T^*\gtrsim 1.22$. In
the case of the HNC theory, the compressibility route is better than
the virial/energy routes for $T^*\lesssim 0.73$ only.

\section{Conclusion}
In this paper we have considered a three-dimensional fluid of
particles interacting via the PS interaction \eqref{1}. This
potential encompasses the ideal gas in the high-temperature limit
($T^*\to\infty$ or $x\to 0$) and the HS fluid in the low-temperature
limit ($T^*\to 0$ or $x\to 1$). However, at finite temperature the
problem becomes much more difficult.  Even  the one-dimensional case
is not exactly solvable \cite{MS06} since there is no \emph{a
priori} limitation to the number of particles that can interact
simultaneously with a given particle.

The diagrams which appear in the density expansions for the PS
fluids are exactly the same as those appearing for HS fluids, except
that each diagram needs to be multiplied by the
temperature-dependent parameter $x$ raised to the number of bonds.
By exploiting this fact, we have obtained the  cavity function
through second order in density and the equation of state through
the fourth virial coefficient. In order to obtain $y_2(r|T^*)$, we
have needed to extend to $r<1$ the functions evaluated by Nijboer
and van Hove \cite{NvH52} for $r>1$. Nevertheless, the possible
analytical evaluation of the elementary-diagram function $\chi(r)$
for $r<1$ seems to be a formidable task. Thus, we have resorted in
that case to two complementary approaches: (i) a numerical
computation by MC integration with an error bar of the order of
$0.001\%$  and (ii) a sixth-degree polynomial approximation
constructed by enforcing seven exact constraints. Both methods show
such an excellent mutual agreement that the results obtained from
the polynomial approximation can be considered as exact from a
practical point of view.

The results obtained here for $y_2(r|T^*)$ and $b_4(T^*)$ have been
compared with those corresponding to the two classical integral
equation theories, namely the HNC and PY theories. It is known that
the PY theory is much better than the HNC one for HS fluids, so that
one could have expected a similar situation for PS fluids, at least
at low temperatures. Our results show that this is indeed the case,
provided that $T^*\lesssim 1$. However, even at very low
temperatures (including the HS limit $T^*\to 0$), the PY solution
strongly underestimates the cavity function in the overlapping
region. This reflects the fact that the fortunate practical
cancelation (in the case of HS) of the diagrams neglected by the PY
equation  does not apply for $r<1$. In this respect, it is
interesting to note that the widely extended belief that the PY
theory becomes exact in the special case of one-dimensional hard
rods is only correct for $r>1$ \cite{MS06}.

When comparing the exact fourth virial coefficient with the HNC and
PY theories one has to take into account their thermodynamic
inconsistency, yielding two HNC predictions (virial/energy and
compressibility routes) and three PY predictions (virial, energy,
and compressibility routes). All these predictions capture the
non-monotonic behavior of  $b_4(T^*)$. In both theories, the
compressibility route is the best one for $T^*\lesssim 0.7$, while
the virial route is preferable if $T^*\gtrsim 0.7$. As in the case
of the structural functions, the equation of state is better
described by the HNC equation than by the PY equation for high
enough temperatures ($T^*\gtrsim 1$).

 It is obvious
that access to non-trivial exact information on the structural and
thermodynamic properties of fluids, even if restricted to special
cases, is of paramount importance. {}From that point of view, we
hope that the results reported in this paper can contribute to an
advancement on our knowledge of the behavior of systems of particles
interacting through bounded potentials.

\acknowledgments

One of the authors (Al.M.) is grateful to the Junta de Extremadura
for supporting his stay at the University of Extremadura in the
period October--December 2005,  when this work was started. His
research has been
 partially supported by the Ministry of Education, Youth, and Sports of
the Czech Republic under the project No.\ LC 512 and by the Grant
Agency of the Czech Republic under project No. 203/06/P432. The
research of the other author (A.S.) has been supported by the
Ministerio de Educaci\'on y Ciencia (Spain) through Grant No.\
FIS2004-01399 (partially financed by FEDER funds) and by the
European Community's Human Potential Programme under contract No.\
HPRN-CT-2002-00307, DYGLAGEMEM.

\appendix*
\section{Evaluation of $\chi''(0)$\label{appA}}
The function $\chi(r)$ is represented by the elementary diagram
displayed at the end of the right-hand side of Eq.\ \eqref{2.4}.
Thus,
\beq
\chi(r_2)=\int d\mathbf{r}_3 \int d\mathbf{r}_4
f(r_3)f(r_4)f(r_{23})f(r_{24})f(r_{34}),
\label{A1}
\eeq
where here $f(r)=f_\hs(r)=-\Theta(1-r)$. Now we differentiate  with
respect to $r_2$ and take into account the mathematical property
\beq
\frac{\partial f(r_{23})}{\partial
r_2}=\delta(r_{23}-1)\frac{\partial r_{23}}{\partial
r_2}=\delta(r_{23}-1)\frac{\mathbf{r}_2\cdot\mathbf{r}_{23}}{r_2}.
\label{A2}
\eeq
The result is
\beqa
\chi'(r_2)&=&2 \int d\mathbf{r}_3 \int d\mathbf{r}_4
f(r_3)f(r_4)f(r_{24})f(r_{34})\nn
&&\times\delta(r_{23}-1)\cos\theta_{23},
\label{A3}
\eeqa
where $\theta_{23}$ is the polar angle of the vector
$\mathbf{r}_{23}$ and the $z$ axis is assumed to point in the
direction of $\mathbf{r}_2$. Making the change of variables
$\mathbf{r}_3\to \mathbf{r}_{23}$, $\mathbf{r}_4\to
\mathbf{r}_{24}$, Eq.\ \eqref{A3} becomes
\beqa
\chi'(r_2) &=&2\int d\mathbf{r}_3 \int d\mathbf{r}_4
f(r_{23})f(r_{24})f(r_{4})f(r_{34})\nn
&&\times\delta(r_{3}-1)\cos\theta_{3}.
\label{A4}
\eeqa
Note that $r_{23}^2=r_2^2+1-2r_2\cos\theta_3$, so that a necessary
(but not sufficient) condition for $f(r_{23})\neq 0$ is
$\theta_3<\pi/2$. Therefore, in the limit $r_2\to 0$, one has
\beq
\chi'(0)=4\pi\gamma(1)\int_0^{\pi/2}d\theta_3\sin\theta_3\cos\theta_3
=\frac{\pi^2}{36} 30,
\label{A5}
\eeq
in agreement with Eq.\ \eqref{4.13}.

Now we differentiate again with respect to $r_2$ to get
\beq
\chi''(r_2)=\chi_1''(r_2)+\chi_2''(r_2),
\label{A6}
\eeq
where
\beqa
\chi''_1(r_2) &=&2\int d\mathbf{r}_3 \int d\mathbf{r}_4
f(r_{24})f(r_{4})f(r_{34})\nn
&&\times\delta(r_{3}-1)\cos\theta_{3}\delta(r_{23}-1)\cos\theta_{23},
\label{A7}
\eeqa
\beqa
\chi''_2(r_2) &=&2\int d\mathbf{r}_3 \int d\mathbf{r}_4
f(r_{23})f(r_{4})f(r_{34})\nn
&&\times\delta(r_{3}-1)\cos\theta_{3}\delta(r_{24}-1)\cos\theta_{24}.
\label{A8}
\eeqa
Let us first consider $\chi_1''(r)$. Note that
$\cos\theta_{23}=r_2-\cos\theta_3$, where it has been taken into
account that $r_{3}=1$. Now, using the property
\beq
\delta(h(x))=|h'(x_0)|^{-1}\delta(x-x_0),
\label{A9}
\eeq
where $h(x)$ is a function that vanishes at $x=x_0$, we have
\beq
\delta(r_{23}-1)=r_2^{-1}\delta(\cos\theta_3-r_2/2).
\label{A10}
\eeq
Thus,
\beq
\chi''_1(r_2) =2r_2\int_0^{2\pi} d\phi_3 \int d\mathbf{r}_4
f(r_{24})f(r_{4})f(r_{34}),
\label{A11}
\eeq
where
$r_{34}^2=r_4^2+1-r_4[r_2\cos\theta_4+\sqrt{4-r_2^2}\sin\theta_4\cos(\phi_3-\phi_4)]$,
$\phi_3$ and $\phi_4$ being azimuthal angles. At the origin one
simply has
\beq
\chi_1''(0)=0.
\label{A12}
\eeq

In Eq.\ \eqref{A8}, since $r_4^2=r_2^2+1-2r_2\cos\theta_{24}$, a
necessary condition for $f(r_4)\neq 0$ is $\theta_{24}<\pi/2$. Now,
setting $r_2=0$ and taking into account that
$\cos\theta_{24}\to-\cos\theta_4$, Eq.\ \eqref{A8} becomes
\beqa
\chi''_2(0)&=&-2\int_0^{2\pi}d\phi_3 \int_0^{2\pi}d\phi_4\int_0^1
d(\cos\theta_3)\cos\theta_3 \nn &&\times\int_{-1}^0
d(\cos\theta_4)\cos\theta_4 f(r_{34}),
\label{A13}
\eeqa
where now $r_{34}^2=2[1-\cos\theta_3
\cos\theta_4-\sin\theta_3\sin\theta_4\cos(\phi_3-\phi_4)]$. The
changes $z=\cos\theta_3$, $z'=-\cos\theta_4$, and
$\phi=\phi_3-\phi_4$ lead to
\beqa
\chi''_2(0)&=&-8\pi\int_0^1dz\int_0^1 dz' z z'\int_0^{\pi}d\phi\nn
&&\times \Theta\left(\cos\phi-\frac{1+2z
z'}{2\sqrt{(1-z^2)(1-{z'}^2)}}\right).
\label{A14}
\eeqa
It can be easily seen that ${1+2z z'}<{2\sqrt{(1-z^2)(1-{z'}^2)}}$
if and only if $z^2+{z'}^2+z z'<3/4$. This requires that
$0<z<\sqrt{3/2}$ and
 $0<z'<(\sqrt{3(1-z^2)}-z)/2$. Consequently,
\beqa
\chi''_2(0)&=&-8\pi\int_0^{\sqrt{3}/2}dz\int_0^{(\sqrt{3(1-z^2)}-z)/2}
dz' z z'\nn&&\times\cos^{-1}\frac{1+2z
z'}{2\sqrt{(1-z^2)(1-{z'}^2)}}.
\label{A15}
\eeqa
The result of the integral is
\beq
\chi''_2(0)=-\frac{\pi^2}{36}\left(12-\frac{18\sqrt{3}}{\pi}\right).
\label{A16}
\eeq

\end{document}